\documentstyle[12pt,graphicx]{article}        
\begin{document}
\baselineskip=12pt
\bibliographystyle{unsrt}

\newcommand{\bmu}{{\bf \mu}}

\begin{center}
 
{\Large{\bf 
 Hydrogen Bond Dynamics Near A Micellar Surface: 
Origin of the Universal Slow Relaxation at Complex Aqueous Interfaces}}\\
\vspace{1cm}
{\large{\bf Sundaram Balasubramanian$^{1}$$^{*}$, Subrata Pal$^{2}$,   
and Biman Bagchi}}$^{2}$$^{*}$\\
\vspace{0.5cm}
$^{1}$ Chemistry and Physics of Materials Unit, \\
Jawaharlal Nehru Centre for Advanced Scientific Research,\\
Jakkur, Bangalore 560064, India.\\
\vspace{0.1cm}
$^{2}$Solid State and Structural Chemistry Unit,
Indian  Institute of Science,\\
Bangalore 560012, India.\\ 
\end{center}
\begin{center}
{\large Abstract}
\end{center}
{\bf The dynamics of hydrogen bonds among water molecules themselves
and with the polar head groups (PHG) at a micellar surface have been investigated
by long molecular dynamics simulations. The lifetime of the hydrogen
bond between a PHG and a water molecule is found  to be much
longer than that between any two water molecules, and is likely to be a general 
feature of hydrophilic surfaces of organized assemblies.
Analyses of individual water trajectories suggest that 
water molecules can
remain {\it bound to the micellar surface for more than a hundred picosecond}.
The activation energy for such a transition from the bound to a free state for the water
molecules is estimated to be about 3.5~kcal/mole.}
\newpage

 The study of hydrogen bond dynamics has proven to be a very useful
tool~\cite{chandler,chandra,bakker} to understand the origin of many
fascinating dynamical properties of water which are due
to its extended hydrogen bond network~\cite{franck}. 
However, the same properties exhibit quite different behavior for water molecules at the surfaces
of self-organized assemblies and biological 
macromolecules~\cite{fleming95,faraday96,lang99,nandi00,zewail02}.
Recent time domain spectroscopic measurements have shown that their dynamics is
considerably slower than their counterparts in bulk water,  sometimes
slower by {\it more than two orders of magnitude}!
While the reorientation of water molecules and solvation of ions or dipoles
in bulk water proceeds with an average time constant of less than a picosecond (ps), the same
at protein surfaces gets extended to hundreds of 
picoseconds~\cite{fukuzaki95,lang99,nandi00,zewail02}. 
Such slow dynamics has been observed in proteins~\cite{lang99,zewail02},
microemulsions~\cite{nandi00,sarkar96}, micelles~\cite{nandi00,pal},
lipid vesicles and bilayers~\cite{poland-cpl}.

 The slow dynamics of water molecules at heterogeneous surfaces seems to be universal and 
could be a collective effect
originating from the surface and/or from  the
nature of the hydrogen bond network at the surface.  
The work of Cheng and Rossky~\cite{rossky98} have already shown that water
molecules at the surface of a protein can be structurally different from
those in the bulk.  A phenomenological theory 
developed recently proposes to explain the emergence of slow dynamics in terms of
a dynamical  equilibrium between bound and free water molecules at the surface~\cite{nandi97}.
Recent atomistic molecular dynamics simulations
of an aqueous micellar assembly, the cesium pentadecaflurooctanoate (CsPFO), 
have shown that  the orientational correlation function of the water molecules
near the micelle show a dramatic slowing down in its long time decay~\cite{bbjpc01}. 
Note that micelles are not only of great interest themselves, but they also 
mimic surfaces of many biological molecules. The presence of a compact hydrophobic core and
a hydrophilic surface resembles proteins rather closely. 

In this study, we explore the two most important microscopic aspects
of water dynamics near the micellar surface -- the hydrogen bond lifetime
of the water-PHG hydrogen bond and also the  {\it distance dependence} of the 
dynamics of the water-water hydrogen bond. The best way
to study this dynamics is to investigate the hydrogen bond time correlation
functions, $S_{\rm HB}(t)$ and $C_{\rm HB}(t)$ (defined below), 
introduced originally by Rapaport~\cite{rapaport} and used  
by Chandler {\it et al}~\cite{chandler} to study pure water and
more recently by 
Chandra~\cite{chandra} to explore effects of ions in water on the lifetime of the 
hydrogen bond.
The time correlation function $S_{\rm HB}^{\rm wPHG}(t)$ for 
the hydrogen bond between 
the polar head group and the water molecules on the surface is defined by
\begin{equation}
S_{\rm HB}^{\rm  wPHG}(t) = \langle h_{\rm wPHG}(0) H_{\rm wPHG}(t) \rangle /\langle h_{\rm wPHG} \rangle
\end{equation}
\noindent where the population variable $h_{\rm wPHG}(t)$ is unity when a particular
tagged hydrogen bond between a water molecule and a polar head group is hydrogen bonded
at time $t$ according to an adopted definition, and zero, otherwise.
On the other hand, 
$H_{\rm wPHG}(t)=1$ if the tagged polar head group-water hydrogen bond remains
{\it continuously} hydrogen bonded during the time duration $t$ and zero otherwise. Thus,
$S_{\rm HB}^{\rm wPHG}(t)$ describes the lifetime of a tagged pair.  The angular brackets denote 
averaging over initial time values and over all water-PHG pairs.  We employed a geometric 
definition of the $\rm wPHG$ hydrogen bond such that a water molecule was assumed to be 
hydrogen bonded to a surfactant if the distance between the oxygen of the water molecule
and the carbon of the headgroup was less than 4.35\AA, and that the oxygen of the water molecule and 
the oxygen of the headgroup was less than 3.5\AA. These distance criteria were obtained from
the first minimum in the respective pair correlation functions. 
For water-water hydrogen bonds,
we have used the same criteria as used by Chandra~\cite{chandra}, which were found to hold 
good for water-water hydrogen bonds near the interface also.

 In order to understand the structural relaxation of a tagged hydrogen bond, one also 
defines the following time correlation function, $C_{\rm HB}^{\rm wPHG}(t)$,
\begin{equation}
C_{\rm HB}^{\rm wPHG}(t) =  \langle h_{\rm wPHG}(0) h_{\rm wPHG}(t) \rangle /\langle h_{\rm wPHG}\rangle
\end{equation}
\noindent Unlike  $S_{\rm HB}^{\rm wPHG}(t)$, this correlation function is expected to exhibit a
longtime tail~\cite{chandler,chandra}. 
 
The surfactant in this simulation is pentadecafluorooctanoate, 
with cesium being the counterion, commonly referred to as CsPFO. The
CsPFO-H$_2$O system has been well studied 
experimentally~\cite{boden7993} and is regarded as a 
typical binary to exhibit micellization.
The amphiphiles form disc shaped micelles, stable over 
an extensive range of concentration and temperature.  
Details of our MD simulations have been discussed 
elsewhere~\cite{bbjpc01,sbbjcp02}. 
The molecular dynamics simulation was carried out in the NVT ensemble
for an aggregate of 62 CsPFO molecules in 10,562 water molecules.
The potential for water
molecules is the extended simple point charge (SPC/E) model~\cite{berendsen87}, 
the  counterions
carry a unit positive charge, which is compensated by a $+0.4e$ charge on 
the carbon of the octanoate headgroup and a $-0.7e$ charge on each of the 
oxygens of the headgroup~\cite{watanabe91}. 
The equations of motion were integrated 
with the RESPA scheme~\cite{tuckerman92} using 
the PINY-MD program~\cite{martyna_unpub} with an outer timestep 
of 4~fs. 
The analyses
reported here were carried out from different sections of a subsequent 3.5~ns trajectory.
The S(t) and C(t) functions were obtained with  time resolutions of 12~fs and 1~ps respectively.
The results reported here are at 
a temperature of 300K.

 Figure 1 shows the time dependence of the correlation function $S_{\rm HB}^{\rm wPHG}(t)$,
with the inset showing the $S_{\rm HB}^{\rm ww}(t)$
(i.e., water-water) correlation function
in bulk water, for comparison. Note  the lengthening of the time scale in
the H-bonding with the PHG of the micelle. $S_{\rm HB}^{\rm wPHG}(t)$ can be fitted to a sum
of three exponentials, with the longest time constant being equal to 9~ps which is
to be compared with 0.97~ps in pure water. The amplitudes
and time constants are given in Table 1~\cite{note1}. From these, one can obtain an 
average time constant of around 6.8~ps, which is thirteen times larger than
its value for the water-water hydrogen bond in pure water. It is difficult to
pinpoint the reason for the slowdown of $S_{\rm HB}(t)$ at the surface. We found that
the water molecules at the surface form bridge hydrogen bonds involving PHGs of the 
nearest neighbor surfactant molecules which could stabilize the $\rm wPHG$ hydrogen
bond. 

 In figure 2, we show the hydrogen bond time correlation function  $C_{\rm HB}^{\rm wPHG}(t)$.
 In the inset we show $C_{\rm HB}^{\rm ww}(t)$ for bulk water. Note again the lengthening of
the long time decay. The values of the time constants
and the respective amplitudes are given in Table 1. In this case, the long time constant 
stretches to more than one hundred picoseconds. The reason for this unusual long decay time can
be traced back to those trajectories which leave the micellar surface to go to  the bulk but
return after a long time to get bonded to the {\it same PHG} 
 at the surface. This is in line with the
explanation of the long time decay of the $C_{\rm HB}(t)$ function in 
bulk water~\cite{chandler,note2}.
In the present problem, the existence of such trajectories indicate
presence of correlations in the surface region. The $S(t)$ function is thus a more accurate
representation of the lifetime dynamics of the hydrogen bond than the $C(t)$ function, although
the latter contains rich information on the correlated pair diffusion of the water molecules.

 Analysis of the trajectories of the individual water molecules at the micellar
surface reveals an amazing richness of events. In figure~3, we show representative trajectories 
over
800~ps, of four arbitrarily chosen water molecules that were hydrogen bonded 
to the surface at time zero. The plot shows the shortest distance, $D_{\rm W-PHG}$, of 
each of the water molecules to the micellar surface as a function of time~\cite{note3}.
For example, the water molecule shown in the top left figure, 
rattles near the surface for about 150~ps, then diffuses into the
bulk region, stays there for around 500~ps, and then revisits the micelle surface, 
presumably to form
a new hydrogen bond with a PHG.
It should also be noted that the micelle itself is quite fluxional over 
these time scales. Note also that while the $S_{\rm HB}^{\rm wPHG}(t)$ 
and $C_{\rm HB}^{\rm wPHG}(t)$ functions study the dynamics of a 
particular wPHG hydrogen bond (with average lifetimes of 6.9 and 43.7ps respectively), the 
apparently longer lifetime of bound water molecules shown in the trajectories in figure~3 
arises from the fact that a water molecule can remain bound to the micellar surface 
even after its hydrogen bond with a particular PHG is broken.

Such dynamical processes involving water molecules affect the way they form hydrogen bond between
themselves. We have explored this aspect by studying the water-water hydrogen bond time correlation
functions, for water molecules that belong to different interfacial layers. 
In figure 4 we show the 
lifetime correlation functions, $C_{\rm HB}^{\rm ww}(t)$ and  $S_{\rm HB}^{\rm ww}(t)$ between two water molecules 
that both exist at several regions away from the surface~\cite{note3}.  This function shows 
a sharp slow down as the surface is approached closely. The time constant
obtained from the $S_{\rm HB}$ function for 
the water molecules {\it within} 6\AA~ from the micellar surface is around 
25-30\% larger than
that for water molecules in the bulk region.
 The time constant for the same water
molecules obtained from the $C_{\rm HB}$ function is about 45\% slower than that for bulk water. Again, the
$S_{\rm HB}$ function decays much faster than the $C_{\rm HB}$ one.

The simulation results can be combined with a theoretical model to obtain useful 
information on the micelle-water interaction. The model assumes a dynamical equilibrium
between the ``bound'' and ``free'' states of water molecules at the micellar 
surface~\cite{nandi97}. The dynamical variable $h(t)$ represents the instantaneous
population of the bound state and the correlation function, $C^{\rm wPHG}_{\rm HB}$(t) gives, under
the regression hypothesis (or linear response theory)~\cite{chandler_statmech}, the 
decay of an initial (t=0) bound state and its formation at a later time.
Therefere, we can set, in the long time,
$C^{\rm wPHG}_{\rm HB}$(t) $\sim$ e$^{-{t\over\tau_{\rm BF}}}$, where $\tau_{B\rm F}$ is the time constant
for the bound to free transition. From Table~1, we find the average time constant 
at 300K to be 43.7ps. We can obtain an activation energy $E_{\rm A}$, for this transition using
transition state theory to express the rate constant ($\frac{1}{\tau_{\rm BF}}$), as,
\begin{equation}
\frac{1}{\tau_{BF}} = \frac{k_BT}{h}e^{-\frac{E_A}{k_BT}}
\end{equation}
where, $k_{\rm B}$ is the Boltzmann constant, $T$ is temperature, and $h$ is the Planck constant. 
$E_{\rm A}$ is then found to be 3.34~kcal/mole. A similar analysis of $C^{\rm wPHG}_{\rm HB}$(t) 
at 350K (data not shown here) yields a time constant of 18.9ps. Using an Arrhenius dependence 
of the time constant with temperature, we find that
the activation energy is 3.50~kcal/mole, in close agreement with the result of the transition
state theory presented above, showing the robustness of the above analysis.

 To conclude, we have carried out large scale atomistic molecular dynamics simulations of an
aqueous micellar system, with emphasis on understanding the
hydrogen bond breaking dynamics at the micellar surface. 
We find the rather surprising result that the hydrogen bond
between the micellar polar head group and a water molecule has a much longer
lifetime -- almost 13 times larger than that in the bulk between two tagged
water molecules. This longer lifetime could originate from the
bridge bonds that the surface water molecules form with the 
PHGs and also from the 
coupling to the micelle which acts as a bath for HB excitations. 
 Another interesting result is that the lifetime of hydrogen bonds between
two tagged water molecules increase only by about 25-30\% as one approaches the
micellar surface. Although this slowdown is not as dramatic as the slowdown in
the orientational relaxation of the water molecules with large residence times near
the interface, it is rather sharp. Analysis of the individual
trajectories reveal that the quasi-bound water at the micellar surface (immobilized by double or 
more  hydrogen bond bridges) may be partly responsible for the slow dynamics. 
The activation energy for the transition from a bound to a free state for the water
molecules on the micellar surface, estimated from both the transition state theory and 
from the temperature dependence of the hydrogen bond lifetime, is about 3.5~kcal/mole.
The agreement between the rather simple model and the temperature dependent simulation 
data is remarkable.
We believe that the results discussed here are generic to organized 
assemblies and biological macromolecules that possess a hydrophilic surface or hydrophilic
pockets that are accessible to water.

\vspace*{0.4cm}
\noindent
{\bf Acknowledgments} \\

 It is a pleasure to thank  Prof. A. Chandra for helpful comments on an
earlier version of this manuscript and for pointing out Ref.~\cite{rapaport}.
The research reported here was
supported in part by grants from the Council of Scientific and
Industrial Research (CSIR) and the Department of Science and Technology
(DST), Government of India. 
\ \\

\clearpage
\begin{table}
\caption{Parameters of multi-exponential fits to the water-PHG hydrogen bond time correlation
functions shown in Figures 1 and 2.}
\begin{center}
\begin{tabular}{|c|c|c|}  \hline
Function & Time constant [ps] & Amplitude (\%) \\ \hline
  & 0.3  & 5.9  \\
$S_{\rm HB}(t)$  & 3.6 &  31.5  \\
  & 9.1  & 62.6  \\ \hline
  & 3.4  & 16  \\
$C_{\rm HB}(t)$  & 29.0 &  63  \\
  & 118.5  & 21  \\ \hline
\end{tabular}
\end{center}
\end{table}
\newpage
\clearpage

\begin{figure}
\includegraphics[scale=0.6,angle=270]{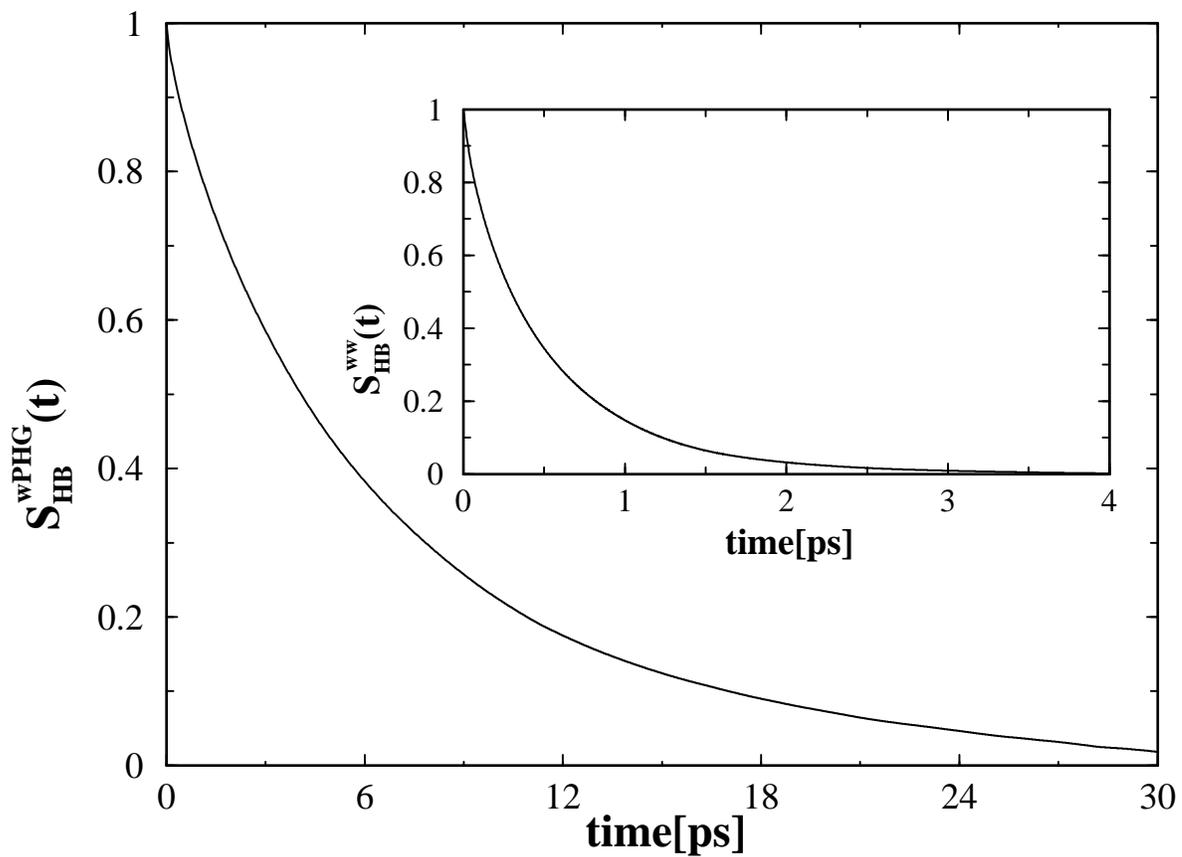}
\caption{S$_{\rm HB}^{\rm wPHG}$(t) function for the
hydrogen bond between the polar head group and water molecules.
Inset shows the same between pairs of water molecules in pure water.}
\end{figure}

\begin{figure}
\includegraphics[scale=0.6,angle=270]{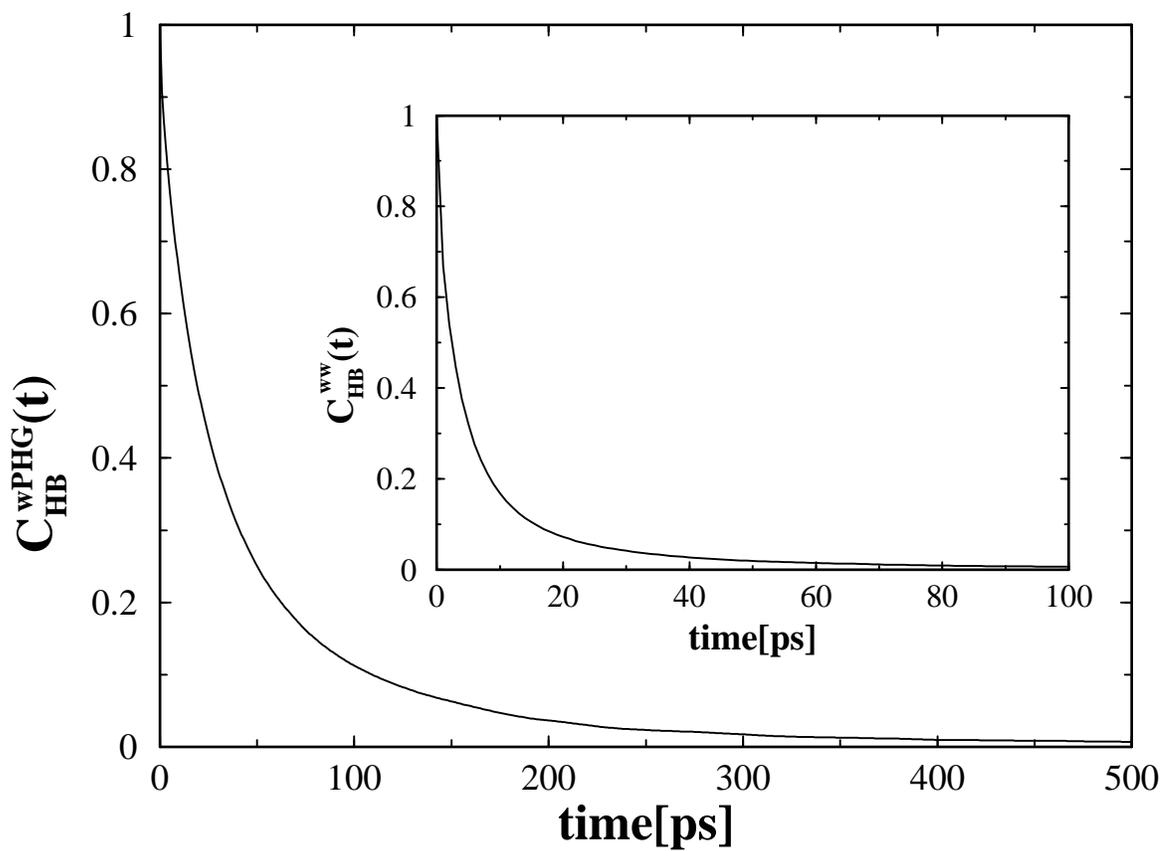}
\caption{C$_{\rm HB}^{\rm wPHG}$(t) function for the hydrogen bond 
between the polar head group and water molecules.
Inset shows the same between pairs of water molecules in the region far away 
 from the micelle~\cite{note4}.}
\end{figure}

\begin{figure}
\includegraphics[scale=0.6,angle=270]{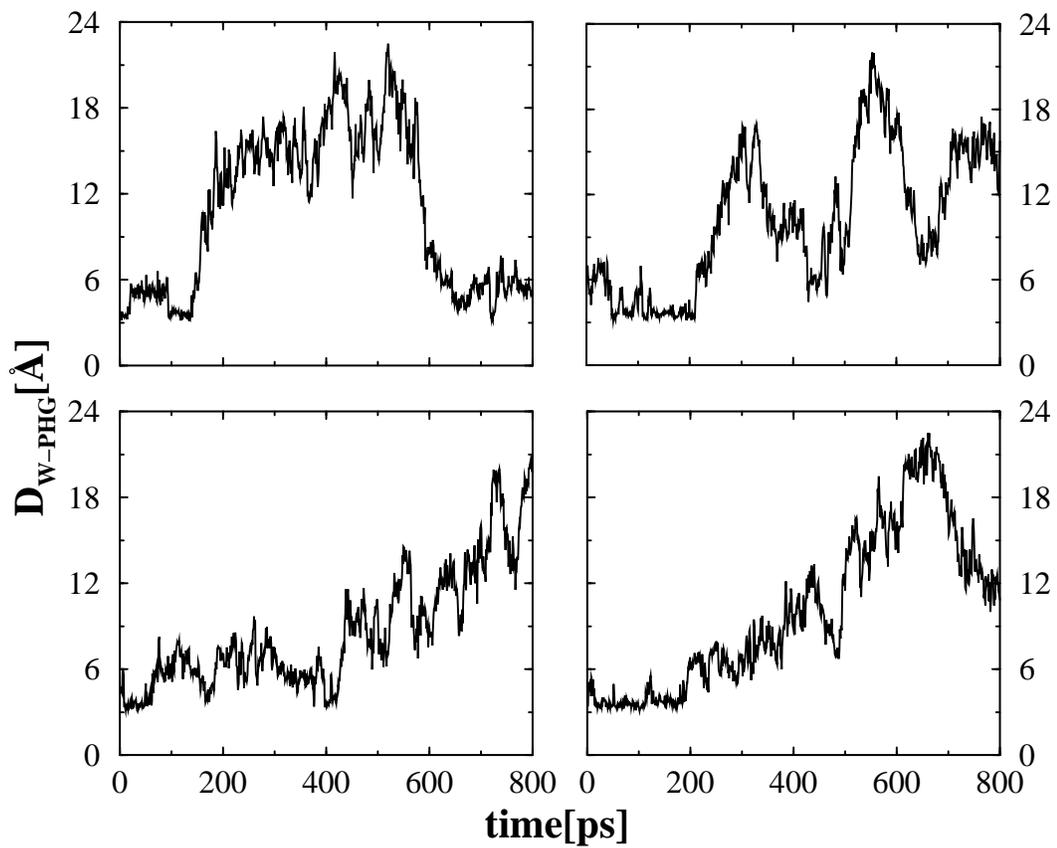}
\caption{Trajectory of four water molecules for
a time period of 800~ps. The shortest distance of the water molecules to the micellar 
surface, $D_{\rm W-PHG}$, is plotted against time.}
\end{figure}

\begin{figure}
\includegraphics[scale=0.6,angle=270]{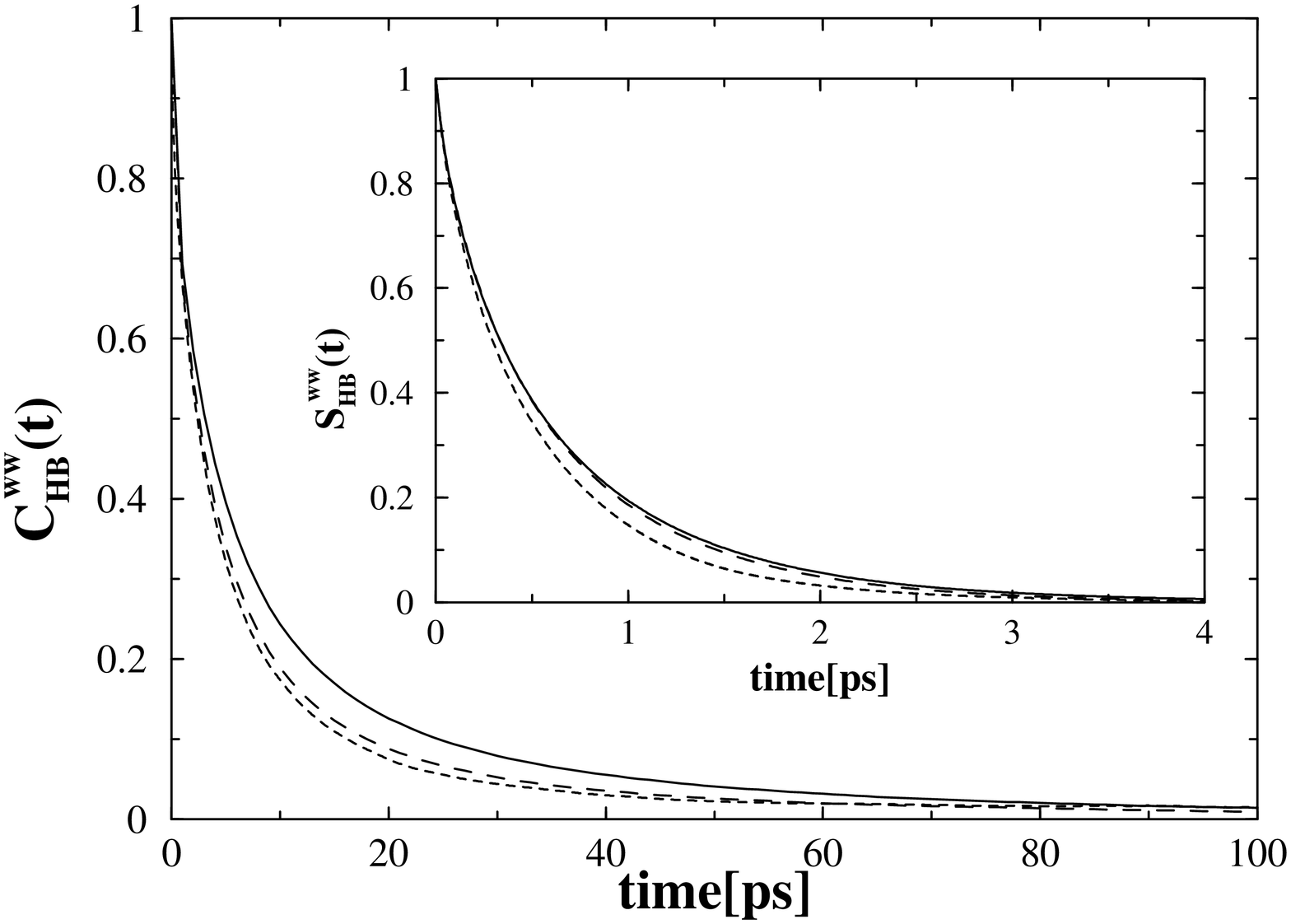}
\caption{The location dependence of the time correlation function,
~C$_{\rm HB}^{\rm ww}$(t), for the hydrogen bond between pairs of water molecules located at different
regions from the micellar surface~\cite{note3}.  
Solid curve: Within 6\AA~ ; Long-Dashed curve: Between 6\AA~ and 9\AA~;
Dashed curve: Beyond 25\AA~.
Inset shows the similar location dependence of the S$_{\rm HB}^{\rm ww}$(t) function for the hydrogen bond 
between pairs of water molecules.}
\end{figure}


\newpage
\clearpage

\begin{thebibliography}{abcd-uf}
\bibitem{chandler} (a) A. Luzar, and D. Chandler, Phys. Rev. Lett. {\bf 76}, 928 (1996);
 Nature (London) {\bf 379}, 53 (1996); 
(b) A. Luzar,  J. Chem. Phys. {\bf 113}, 10663 (2000).

\bibitem{chandra} A. Chandra,  Phys. Rev. Lett. {\bf 85}, 768 (2000).

\bibitem{bakker} M.F. Kropman, and H. Bakker,  Science, {\bf 291}, 2118 (2001);
 H. Xu, and B. Berne, J. Phys. Chem. B, {\bf 105}, 11930 (2001);  
S. Raugei, and M.L. Klein,  J. Am. Chem. Soc., {\bf 123}, 9484 (2001);

\bibitem{franck} F. Francks, Ed.  {\em Water, A Comprehensive Treatise}, (Plenum
Press, New York, 1972-1982);  F. Sciortino {\em et al.} Phys. Rev. Lett. {\bf 64},
1686 (1990).

\bibitem{fleming95} S. Vajda  {\em et al.} J. Chem. Soc. Farad. Trans., {\bf 91}, 867 (1995).

\bibitem{faraday96} See for example,{\it  Hydration Processes in Biological and
Macromolecular systems, Faraday Disc}. 1996, {\bf 103}, 1-394.

\bibitem{lang99}  X.J. Jordanides {\em et al.} J. Phys. Chem. B. {\bf 103}, 7995 (1999).

\bibitem{nandi00} N. Nandi,  K. Bhattacharyya, and B. Bagchi,  Chem. Rev. {\bf 100}, 2013 (2000).

\bibitem{zewail02} S.K. Pal, J. Peon, A.H. Zewail, Proc. Nat. Acad. Sci. {\bf 99}, 1763 (2002).

\bibitem{fukuzaki95}  M. Fukuzaki {\em et al.} J. Phys. Chem. {\bf 99}, 431 (1995); 
G. Otting, in {\it Biological Magnetic Resonance},
 (ed. N. Ramakrishna, L.J. Berliner), KluwerAcademic/Plenum, New York, 1999, vol. 17,
pp. 485. 

\bibitem{sarkar96}  N. Sarkar {\em et al.}  J. Phys. Chem.
{\bf 100}, 15483 (1996).

\bibitem{pal} S.K. Pal {\em et al.} Chem. Phys. Lett. {\bf 327}, 91 (2000); 
R.E. Riter,  D.M. Willard,  and N.E. Levinger,  J. Phys. Chem. B,
 {\bf 102}, 2705 (1998).

\bibitem{poland-cpl} T. Rog, K. Murzyn, and M. Pasenkiewicz-Gierula,  Chem. 
Phys. Lett., {\bf 352}, 323 (2002).

\bibitem{rossky98}Y.-K. Cheng and P.J. Rossky,  Nature, {\bf 392}, 696 (1998).

\bibitem{nandi97} N. Nandi, and B. Bagchi,  J. Phys. Chem. B.  {\bf 101}, 10954 (1997). 

\bibitem{bbjpc01} S. Balasubramanian,  and B. Bagchi,  J. Phys. Chem. B {\bf 105}, 12529,
(2001); {\em ibid} {\bf 106}, 3668 (2002).

\bibitem{rapaport} D.C. Rapaport,  Mol. Phys., {\bf 50}, 1151 (1983).

\bibitem{boden7993} N. Boden,  K.W. Jolley, and  M.H. Smith,  J. Phys. Chem. {\bf 97}, 7678 (1993); H. Iijima {\em et al.}   J. Phys. Chem. B {\bf 102}, 990 (1998).

\bibitem{sbbjcp02}S. Balasubramanian, S. Pal, and B. Bagchi, Curr. Sci. {\bf 82}, 845 (2002);
 S. Pal, S. Balasubramanian, and B. Bagchi, J. Chem. Phys. 
{\bf 117}, 2852 (2002).

\bibitem{berendsen87} H.J.C. Berendsen, J.R. Grigera, and T.P. Straatsma, 
 J. Phys. Chem.  {\bf 91}, 6269 (1987).

\bibitem{watanabe91} K. Watanabe, and M.L. Klein,   J. Phys. Chem. {\bf 95}, 4158 (1991).

\bibitem{tuckerman92} M.E. Tuckerman, B.J. Berne,  and G.J. Martyna,  J. Chem.
Phys. {\bf 97}, 1990 (1992).

\bibitem{martyna_unpub} M.E. Tuckerman {\em et al.} Comp. Phys. Comm.  {\bf 128}, 333 (2000).

\bibitem{note1}  Note that even in bulk water,
the lifetime correlation function $S_{\rm HB}^{\rm ww}(t)$ 
is non-exponential, with three time constants,
0.07 ps (14\%), 0.49 ps (70\%) and 0.97 ps (16\%). The
average lifetime is 0.51 ps.

\bibitem{note2} $C_{\rm HB}^{\rm ww}(t)$ is also non-exponential in bulk
water, with time constants 3.1 ps (61\%) and 17.25 ps (39\%). The average
time constant is 7.2 ps.

\bibitem{note3} The instantaneous positions of the carbon atoms of the headgroups are
taken to denote the micellar surface.

\bibitem{chandler_statmech} D. Chandler, {\em Introduction to Modern Statistical Mechanics},
(Oxford Unversity Press, New York, 1987, p. 242).

\bibitem{note4} The $C_{\rm HB}^{\rm ww}(t)$ function for pure water can show a spuriously long 
time tail which denotes the re-formation of the hydrogen bond of a given pair. This 
effect should vanish in the thermodynamic limit, and the time correlation function will 
not possess a long time tail. The function in the inset for water molecules in the 
far region matches the correlation function obtained for a collection
of 256 water molecules in bulk, exactly, upto 80~ps. 

\end{thebibliography}
\end{document}